\documentclass[conference]{IEEEtran}
\IEEEoverridecommandlockouts
\usepackage{cite}
\usepackage{amsmath,amssymb,amsfonts}
\usepackage{algorithmic}
\usepackage{graphicx}
\usepackage{textcomp}
\usepackage{xcolor}
\usepackage{url}

\usepackage{tabularx}

\usepackage{booktabs}
\usepackage{eurosym}%

\def\BibTeX{{\rm B\kern-.05em{\sc i\kern-.025em b}\kern-.08em
    T\kern-.1667em\lower.7ex\hbox{E}\kern-.125emX}}
\begin{document}

\title{Explaining Deep Learning Embeddings for Speech Emotion Recognition by Predicting Interpretable Acoustic Features\\
\thanks{DML was supported by a RallyPoint Fellowship and NIMH [5T32MH125815-03] and NIDCD training grants [5T32DC000038-28]. SSG was supported by a gift to the McGovern Institute for Brain Research at MIT and by the NIH Bridge2AI program. This work has been submitted to the IEEE for possible publication. Copyright may be transferred without notice, after which this version may no longer be accessible.}}


\author{
\IEEEauthorblockN{Satvik Dixit}
\IEEEauthorblockA{\textit{Dept. of Electrical and Computer Engineering} \\
\textit{Carnegie Mellon University}\\
Pittsburgh, USA \\
satvikdixit@cmu.edu}
\and
\IEEEauthorblockN{Daniel M. Low}
\IEEEauthorblockA{\textit{Harvard Chan School of Public Health} \\
\textit{Harvard University}\\
Boston, USA \\
dlow@g.harvard.edu}
\and
\IEEEauthorblockN{Gasser Elbanna}
\IEEEauthorblockA{\textit{Harvard Medical School} \\
\textit{Harvard University}\\
Boston, USA \\
gelbanna@mit.edu}
\and
\IEEEauthorblockN{Fabio Catania}
\IEEEauthorblockA{\textit{McGovern Institute} \\
\textit{Massachusetts Institute of Technology (MIT)}\\
Cambridge, USA \\
fabiocat@mit.edu}
\and
\IEEEauthorblockN{Satrajit S. Ghosh}
\IEEEauthorblockA{\textit{McGovern Institute} \\
\textit{Massachusetts Institute of Technology (MIT)}\\
Cambridge, USA \\
satra@mit.edu}
}

\maketitle

\begin{abstract}
Pre-trained deep learning embeddings have consistently shown superior performance over handcrafted acoustic features in speech emotion recognition (SER). However, unlike acoustic features with clear physical meaning, these embeddings lack clear interpretability. Explaining these embeddings is crucial for building trust in healthcare and security applications and advancing the scientific understanding of the acoustic information that is encoded in them.
This paper proposes a modified probing approach to explain deep learning embeddings in the SER space. We predict interpretable acoustic features (e.g., f0, loudness) from (i) the complete set of embeddings and (ii) a subset of the embedding dimensions identified as most important for predicting each emotion. If the subset of the most important dimensions better predicts a given emotion than all dimensions and also predicts specific acoustic features more accurately, we infer those acoustic features are important for the embedding model for the given task. We conducted experiments using the WavLM embeddings and eGeMAPS acoustic features as audio representations, applying our method to the RAVDESS and SAVEE emotional speech datasets. Based on this evaluation, we demonstrate that Energy, Frequency, Spectral, and Temporal categories of acoustic features provide diminishing information to SER in that order, demonstrating the utility of the probing classifier method to relate embeddings to interpretable acoustic features.
\end{abstract}


\section{Introduction}\label{sec:intro}

Speech Emotion Recognition (SER) involves automatically identifying emotional states from spoken language \cite{10.1145/3129340} and is an important task in several fields, including human-computer interaction \cite{schuller2018age} and mental health assessments \cite{ding2023speech}. While conventional methods rely on handcrafted features, recent breakthroughs in this domain have come from deep neural networks, especially those trained in a self-supervised manner \cite{10.1145/3129340}. These networks learn speech embeddings that outperform traditional features in terms of SER accuracy \cite{turian2022hear, yang2021superb}, but the mechanisms behind their success remain unclear. In particular, the question of what kind of acoustic information deep learning (DL) models use for a particular task remains largely unanswered \cite{mohamed2022self}.

This paper focuses on explainability by using \textit{probing classifiers} to investigate the acoustic information contained in DL embeddings.
Using a novel tiered prediction strategy, we aim to identify the specific interpretable acoustic feature information that is more relevant for distinguishing emotions using DL embeddings. This can enhance our understanding of the mechanisms driving the success of DL models in SER.


\begin{figure*}[htb]
\centering
\includegraphics[width=1\textwidth]{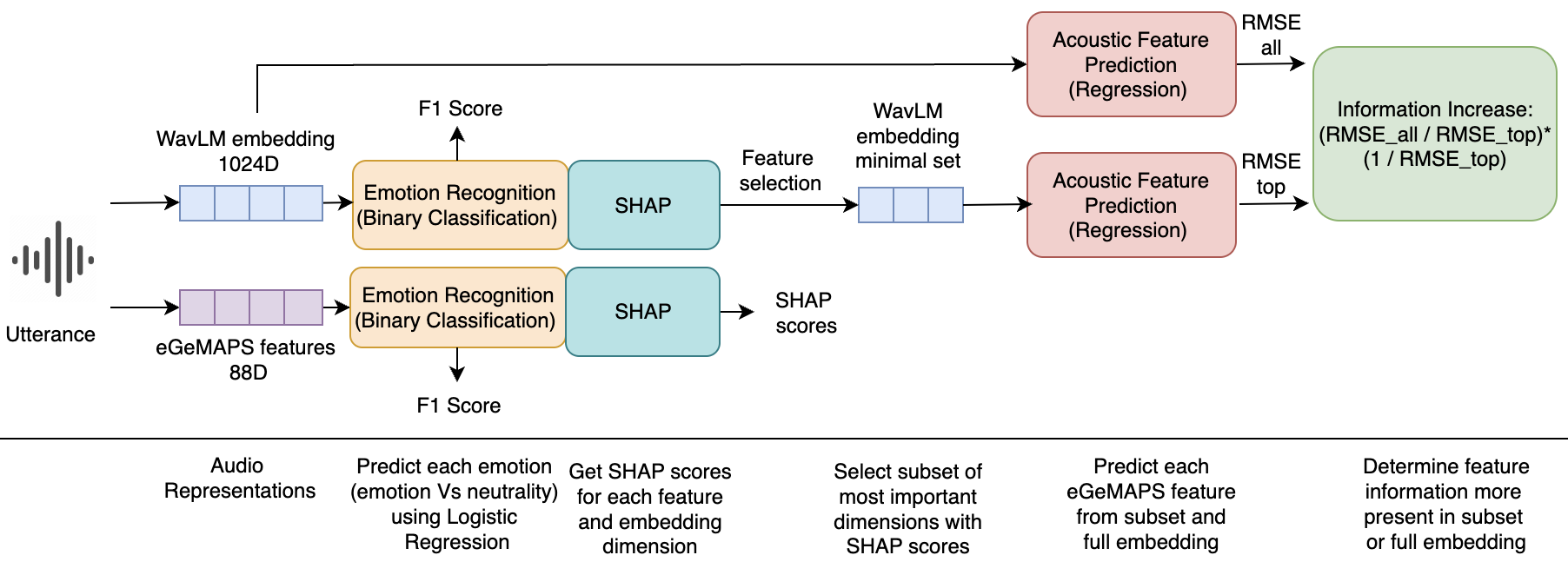}
\caption{The training and testing pipeline for our SER models and the top vs. all dimensions information increase method.}
\label{fig:architecture}
\end{figure*}

In this study, our contributions are that we first provide insights into what types of acoustic features characterize different emotions, using the standard eGeMAPS feature set in a purely interpretable model. Second, we quantify how well these interpretable features are predicted from WavLM DL embeddings, offering insights into the information contained by these embeddings. Our methodological contribution is that we demonstrate which type of acoustic features are better represented in the subset of embedding dimensions that most characterize a given emotion and provide a new metric, information increase, to quantify this. We hypothesize that these acoustic features are relevant to that emotion. Our primary focus is explainability: gaining an understanding of the information encoded in these models can support their further improvement and their applications in various tasks and bolster user trust in the predictions. A tutorial is made available 
\footnote{Our code and tutorials are made publicly available at \url{https://github.com/satvik-dixit/explainability_SER }} and could be used to probe different self-supervised embeddings on different acoustic features in any audio-related classification task.

\section{Related Work}
\label{sec:related work}

Probing classifiers have been widely used to analyze text embeddings \cite{belinkov2021probing}. However, less work has been done in the audio domain. One recent study probed transformer-based audio models for emotion recognition content to understand how much information related to emotions is contained in different models and layers \cite{2023arXiv230808713S}, but did not probe for specific acoustic information. Another study fine-tuned pre-trained models to detect emotional properties (a multitask output: arousal, valence, and dominance) \cite{Triantafyllopoulos_2022}. They then probed these models for a set of acoustic features, comparing a pre-trained Wav2Vec 2.0 \cite{baevski2020wav2vec20frameworkselfsupervised} model fine-tuned with an added output head versus additionally fine-tuning the transformer layers. If a feature is represented more effectively after fine-tuning the transformer layers, resulting in improved predictions of acoustic features, then it is hypothesized that this information is encoded in the model or captured by the model. Since fine-tuning these layers improved performance, more information about certain acoustic features would indicate that they are relevant to the task. However, they did not find changes in information except that audio duration information became less important for the improved model. Another recent study compares Wav2Vec 2.0 representations with selected eGeMAPS features \cite{li2022explorationselfsupervisedspeechmodel}, however they used canonical correlation analysis instead of probing.



\section{Methods}
\label{sec:format}

\subsection{Datasets}
\label{sec:dataset}

We selected two well-established emotional speech datasets to evaluate how outcomes vary between them: the Ryerson Audio-Visual Database of Emotional Speech and Song (RAVDESS) \cite{livingstone2018ryerson} and the Surrey Audio-Visual Expressed Emotion (SAVEE) \cite{haq2011multimodal}. The choice of these datasets was motivated by their numerous similarities, which allow for a direct and meaningful comparison of different feature extraction techniques. Both datasets are in English. Also, they are balanced in actors' gender, spoken sentences, and expressed emotions (\textit{the Big Six} \cite{ekman1992argument} plus neutrality), minimizing potential biases due to data imbalance. Additionally, they were recorded in controlled laboratory environments, ensuring high-quality, noise-free audio samples, which can reduce bias during feature extraction and training.

\subsection{Audio representations}
\label{ssec:embeddings}
We looked at two categories of audio representations:

\textit{Handcrafted acoustic features}: These are interpretable features designed to capture specific aspects of the audio signal, such as intensity, frequency, spectral, and temporal elements. For this study, we used eGeMAPS, a widely adopted standard set of acoustic features (implemented using OpenSMILE eGeMAPSv02 \cite{eyben2010opensmile}), which has proven to be somewhat effective for emotion recognition \cite{eyben2015geneva}.

\textit{Deep learning embeddings}: These are representations learned through neural networks that can capture complex and abstract patterns in the audio signal. In this study, we used WavLM Large \cite{Chen_2022}, a pre-trained speech self-supervised model that has demonstrated state-of-the-art performance on the SUPERB benchmark for emotion recognition \cite{yang2021superb}. The embeddings were mean-pooled over time to get one embedding per utterance.

\subsection{SER classification using eGeMAPS and WavLM}
\label{ssec:SER}
As displayed in Figure \ref{fig:architecture}, we employ a binary classifier for each distinct emotion category (emotion vs neutrality). Specifically, we divide the dataset such that half of the samples represent one particular emotion, while the remaining half are utterances of the neutral emotion. This is done so we can focus on the features that characterize an emotion versus neutral, which we find simpler to interpret than what characterizes an emotion versus other emotions (e.g., anger should be louder than neutral). 

Prior to classification, we perform speaker normalization. For the classification task, we employ Logistic Regression with L2 regularization - this helps reduce collinearity issues \cite{schreiber2018ridge}. We perform hyperparameter tuning on the regularization parameter or `C' with the values from the set \{0.01, 0.1, 1, 10, 100\} using use 5-fold nested cross-validation (i.e., the optimal parameter is chosen within the training set). Samples of a specific speaker are either in the training or test set, not both. We use F1 score to assess classifier performance.

\subsection{Determining top eGeMAPS Features and WavLM Embedding Dimensions using SHAP}
\label{ssec:FIegemaps}
\label{ssec:FIHybridBYOLs}

We rank the eGeMAPS features in order of importance for SER classification using SHAP \cite{NIPS2017_7062} in order to determine the most important features and feature categories for predicting each emotion, albeit in a less optimal, but interpretable model. The eGeMAPS feature categories \cite{eyben2015geneva} are described in Table \ref{tab:egemaps_categories}.

We also find the most important dimensions for the WavLM DL model for each emotion in the same way to rank each of the 1024 WavLM embedding dimensions in order of importance. 
We perform a post-hoc analysis to determine the minimal set of most important dimensions by taking the lowest number of features at which the performance is the highest in classifying each emotion (also with Logistic Regression with L2 regularization). To determine this set for the WavLM embedding dimensions, we sweep the feature importance vector in steps of 10 starting from the 10th most important feature. This number for the estimated minimal set for each emotion is reported in Table \ref{tab:performance}. Next we described our probing approach to better understand the information encoded in these subsets of embedding dimensions.

\begin{table}[ht]
\caption{eGeMAPS feature categories}
\begin{tabular}{p{0.35\linewidth} | p{0.6\linewidth}}
\toprule
Category & eGeMAPS Features \\
\midrule
Energy / Amplitude & loudness; sound Level; HNR; shimmer\\
\midrule
Frequency & F0; Jitter;  Formants 1–3 frequency and bandwidth \\
\midrule
Spectral balance & Alpha ratio; Hammarberg index; Spectral slope; MFCCs 1-4; H1-H2 and H1-A3; Spectral flux\\
\midrule
Temporal & Rate of loudness peaks; Mean length and SD of Voiced and unvoiced segments; syllable rate \\
\bottomrule
\end{tabular}
\label{tab:egemaps_categories}
\end{table}

\subsection{Probing handcrafted eGeMAPS Features from Top WavLM Embedding Dimensions}
\label{ssec:prediction}

\begin{figure*}[htb]
\centering
\includegraphics[width=1\textwidth]{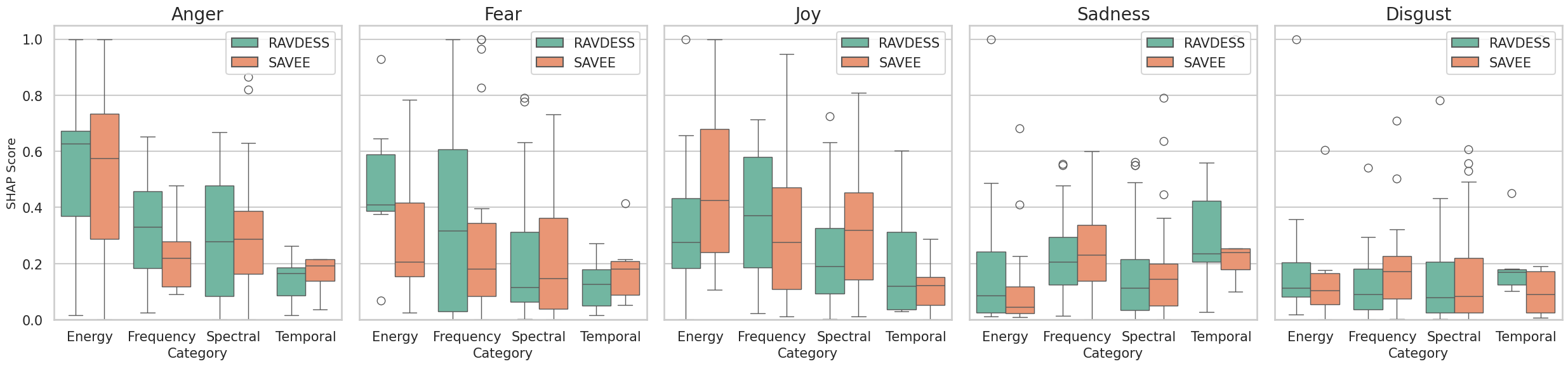}
\caption{Normalised SHAP scores to show feature importance of eGeMAPS feature categories in interpretable model}
\label{fig:res}
\end{figure*}

\begin{figure*}[htb]
\centering
\includegraphics[width=1\textwidth]{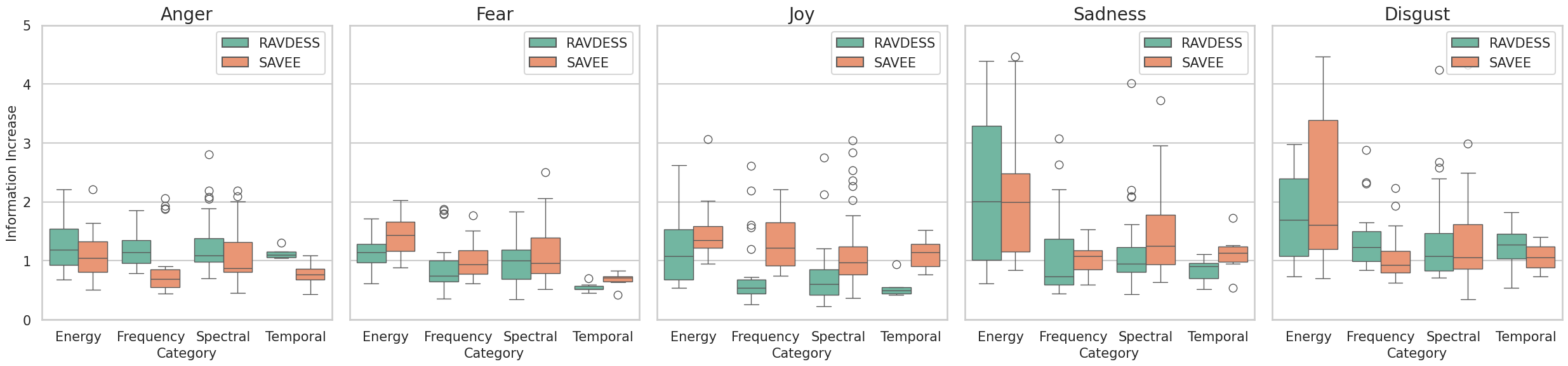}
\caption{
Average information increase per eGeMAPS feature category for all emotions
}
\label{fig:info_increase}
\end{figure*}
\label{sec:IG_wide}

We estimate how much of each acoustic feature is contained in the WavLM embeddings by the ability of the embedding to predict the feature. We train our model on the minimal subset of top dimensions to predict the eGeMAPS features one at a time. For prediction, we use a Ridge regression model and do hyperparameter tuning on the regularization strength coefficient or `alpha' with the values \{0.001, 0.01, 0.1, 1, 10, 100\}. We only used a linear classification model, as is common in probing classification, to avoid allowing more flexible models to infer new features as we wish to only look at information in the DL embeddings \cite{belinkov2021probing}. 
For every feature, we compute the information increase between all WavLM embedding dimensions and top WavLM embedding dimensions weighted by how well the feature is encoded in the minimum subset (to avoid highlighting features that are not encoded well) using the following custom metric:

\begin{equation}
    \text{Information increase} = \frac{RMSE_{all}}{RMSE_{top}}\times\frac{1}{RMSE_{top}}
\end{equation}

Here $RMSE_{all}$ and $RMSE_{top}$ are the RMSE for predicting the given acoustic feature using all dimensions and top dimensions of the WavLM embedding.
We are trying to identify handcrafted features which are encoded much better in the top dimensions of the embedding compared to all dimensions and therefore have a high value of $\frac{RMSE_{all}}{RMSE_{top}}$. These features should also be encoded significantly well in the top dimensions of the embedding (have low error); to enforce this, we add a $\frac{1}{RMSE_{top}}$ term which weighs the score down if the prediction by the top dimensions has a large error. This would indicate that the feature is not well captured by the top dimensions, even if they are better than using all dimensions.

\section{Results}

The DL-based embeddings outperform the handcrafted features for every emotion for both datasets in terms of F1 scores (see Table \ref{tab:performance}), which justifies using the DL models. 

\begin{table}[h!]
\caption{
Comparison of SER F1 scores using eGeMAPS and all WavLM embedding dimensions (WL all) for RAVDESS / SAVEE. The number of top dimensions in the minimal subset is shown in the last column.
}

\begin{tabularx}{\linewidth}{Xccccc}
\toprule
Emotions & eGeMAPS & WL all & \# top dim.\\
\midrule
Anger & 93.1 / 99.4 &       97.7 / 99.4  & 20 / 60\\
Fear & 92.6 / 97.9 &       97.5 / 99.4 &                 30 / 50 \\
Joy & 88.8 / 98.7 &       96.8 / 99.4 &                70 / 60\\
Sadness & 79.0 / 70.0 &      83.0 / 97.5 &               130 / 80 \\
Disgust & 89.1 / 62.4 &       99.1 / 98.8 &               20 / 70 \\
\bottomrule
\end{tabularx}
\label{tab:performance}
\end{table}

Next, we determine top eGeMAPS features and WavLM embedding dimensions using SHAP. Figure \ref{fig:res} shows the most important eGeMAPS feature categories for every emotion. There is variability as to which categories are most important for each emotion (e.g., energy for anger; temporal for disgust). To compare this to the information used by WavLM model, we probe handcrafted eGeMAPS features from the subset of top WavLM embedding dimensions. For every emotion, we compute the average information increase to show the eGeMAPS feature categories that are most relevant to the task as shown in Figure \ref{fig:info_increase}. In general, the median information increase follows the order: Energy \textgreater{} Frequency \textgreater{} Spectral \textgreater{} Temporal for all emotions for both datasets. 

\section{Discussion}
\label{sec:discussion}


By providing a novel probing method and metric, we demonstrate how to estimate interpretable acoustic information contained in DL-based embeddings. Our method helped us find that energy-based features have the most information increase across datasets across almost all emotions. Therefore, we hypothesize that the WavLM embeddings use this information to better perform the SER task. 
Even though the SHAP scores suggest different feature categories are important for every emotion, the energy based features always have the highest median information gain. The lower information increase for temporal features is consistent with the time-pooled nature of the embeddings, suggesting that some time-dependent information may be lost in the process. Furthermore, this method can help us understand why a given emotion is detected better by higher-performing DL-based models. For instance, the embeddings seem to use energy features to classify sadness more than other feature categories, but energy is similarly important to other categories in the handcrafted model, which may explain its lower performance as shown in Table \ref{tab:performance}. More generally, we see an ordering of feature categories that correlates with performance; when this fails in eGeMAPS models for sad and disgust, performance drops. Overall, we can leverage methods that compare models or embeddings that are trained on the same task, but one performs better than the other, and we can compare their relative information content.

A limitation of this method is that while it estimates what information from eGeMAPS is and is not encoded in DL-based embeddings, it does not imply that this is the most important information the DL-based embeddings is using for classification; it might be using other information beyond the eGeMAPS features we tested. However, given a black-box, providing even part of the information encoded using our method can improve explainability. 

An important future direction is investigating the generalizability of our results across more datasets and languages (including those with recordings of naturalistic emotion production such as MSP \cite{lotfian2017building}). The analysis of other handcrafted features or DL-based embeddings could also be explored.

\newpage

\bibliographystyle{IEEEtran}
{\large\bibliography{refs}}
\end{document}